# Organic ferroelectric Croconic Acid: A concise survey from bulk single crystals to thin films

*Sambit Mohapatra,[a*] Salia Cherifi-Hertel,[a] Senthil Kumar Kuppusamy,[a] Guy Schmerber,[a] Jacek Arabski,[a] Benoît Gobaut,[a] Wolfgang Weber,[a] Martin Bowen,[a] Victor Da Costa,[a] and Samy Boukari[a1]*

[a] *Université de Strasbourg, CNRS, Institut de Physique et Chimie des Matériaux de Strasbourg, UMR 7504, F-67000 Strasbourg, France*

**Abstract:** Owing to prospective energy-efficient and environmentally benign applications, organic ferroelectric materials are useful and necessary alternative to inorganic ferroelectrics. Although the first discovered ferroelectric, Rochelle salt, was a salt of an organic compound, organic ferroelectrics have not been as abundant as the inorganic ones. Further, the small polarization values in the organic systems discovered so far have been a demotivating factor for their applications. However, scientific interest and activities surrounding such materials, for the purpose of fundamental understanding and practical applications, have significantly risen lately, especially after the discovery of above-room-temperature ferroelectricity in croconic acid (4,5-dihydroxy-4-cyclopentene-1,2,3-trione, $H_2C_5O_5$) crystals with polarization values rivalling those found in inorganic ferroelectrics. Its large polarization, organic nature, and vacuum sublimability make croconic acid an ideal candidate for non-toxic and lead-free device applications. In this review article, we survey the scientific activities carried out so far involving ferroelectricity in this novel material, paying equal attention to its bulk single crystal and thin film forms. While we discuss about the origin of ferroelectric order and the reversal of polarization in the bulk form, we also summarize the directions toward applications of the thin films.

1. Introduction

As a very intriguing phenomena in condensed matter science, ferroelectricity continues to intrigue researchers several decades after its discovery, with continuously unraveled novel aspects of the phenomenon and with prospects for applications toward solutions of real-world problems in mind. With developments in the technology for material synthesis and characterization, it has been possible to explore this phenomenon at its fundamental level. In parallel, fabrication of ferroelectrics-based electronic devices pertaining to a wide range of fields—for example, sensors and actuators, memory devices, nanoelectronics and spintronics, flexible electronics, bio-compatible electronics, and energy harvesting—has proven the importance of this class of materials for scientific and industrial growth.[1–7] More recently, discoveries such as negative capacitance devices,[8] domain wall memory devices and diodes,[9] polar skyrmions,[10] topologically exotic domain structures and domain walls,[11] and memristor like devices[12] hint at some of the exotic possible future applications that ferroelectric materials may offer.[13]

Owing to the widespread present and future applications that utilize ferroelectrics, it is imperative to address and deal with the issues and limitations that exist with the currently available repository of inorganic ferroelectric materials, and work toward designing novel ferroelectric materials to overcome the challenges. The major concern with the use of inorganic ferroelectrics is related to the presence of toxic elements—lead (Pb), bismuth (Bi) and others. For example, in the most commonly used inorganic ferroelectrics such as lead zirconate titanate ($Pb[Zr_xTi_{1-x}]O_3$ ($0 \leq x \leq 1$)) and lead titanate ($PbTiO_3$), a significant percentage of Pb and other toxic elements are present. Consequently, the increase in global environmental awareness has been encouraging the search for lead-free alternatives.[14–16] This has motivated researchers to look beyond the existing repository and explore newer ferroelectric materials. On one hand, while alternatives such as alkaline niobates ($XNbO_3$, X=Alkali element), barium titanate ($BaTiO_3$)-based compositions,[17–19] and hybrid organic-inorganic structures[20],[21] are being explored; on the other hand, purely organic ferroelectric materials clearly race ahead when it comes to the issue of environmental compatibility.

---

* Corresponding author

Besides environmental suitability, there is a multitude of other aspects by which organic ferroelectrics may provide distinct advantages over the inorganic ones. For example, while their simpler and cheaper fabrication and processing may provide easier commercial manufacturing, their organic nature may be useful to design highly energy efficient electronic devices.[22,23] Further, their lightweight, flexible,[24],[25] and bio-compatible nature promises applications in flexible and bioelectronics devices.[26],[27] Furthermore, the possibility for chemical functionalization can be utilized to enhance properties, bring about newer properties, and design ferroelectric systems targeting specific applications.[28],[29],[30] Despite all the above advantages, the bottleneck with organic ferroelectrics is the scarcity of stable organic systems with sizeable room-temperature (RT) spontaneous ferroelectric polarization. The repository of organic ferroelectrics is not only much smaller compared to the inorganic one, but also only a few of them are ferroelectric at room-temperature or RT. Thus, the need for organic ferroelectrics with higher room-temperature spontaneous polarization, than that of the currently available ones, is leading to

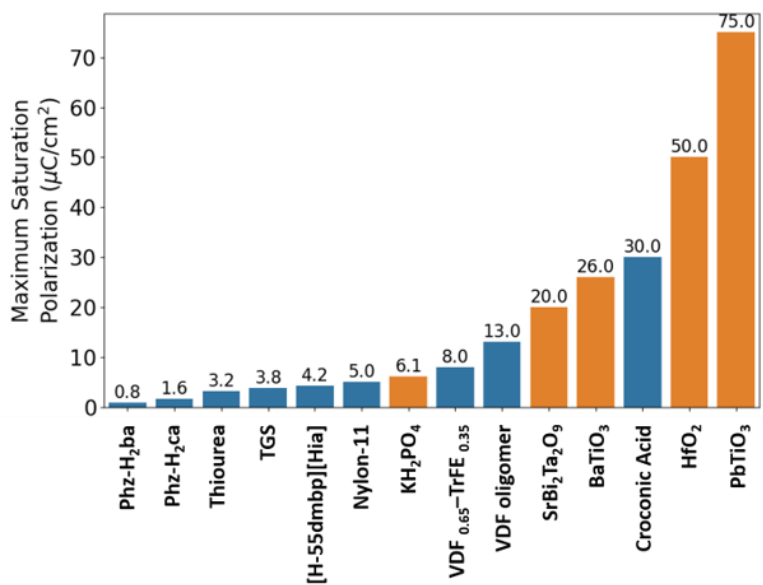

**Figure 1.** Bar chart of maximum saturation polarization of common ferroelectric materials with blue and orange colors representing the organic and the inorganic ones, respectively. Conventional organic ferroelectric materials lie at the lower side of the spectrum, while croconic acid takes a step ahead into the inorganic ferroelectric range.

the synthesis and discovery of new organic systems whose ferroelectricity can be based on different mechanisms, such as charge-transfer,[31],[32],[33] hydrogen bonding,[34–38] and supramolecular assembly.[39],[40],[41]

Among all the organic ferroelectric materials discovered so far, a unique single-component hydrogen bonded molecular system named Croconic Acid— (4,5-dihydroxy-4-cyclopentene-1,2,3-trione, $H_2C_5O_5$); hereafter referred to as CA—stands out. In CA, the polarization reversal takes place via resonance-assisted concerted intermolecular proton transfer and concurrent π-bond relocation. Crystals of CA show a large value of room-temperature polarization of ~30 $\mu C/cm^2$. This value is the highest among all organic ferroelectric materials, and is higher than that for many standard inorganic ferroelectrics as well.[42–44] While several other organic ferroelectric materials are ferroelectric at low temperatures, the presence of room-temperature ferroelectricity in CA makes it a suitable candidate for a wide variety of practical applications. Besides, ferroelectricity in CA is robust even at high temperatures up to 450 K (decomposition temperature) and at high pressures up to 5.3 GPa.[45] Furthermore, ultra-high vacuum (UHV) evaporation of CA makes it possible to fabricate high-quality thin films,[46],[47] which then allows for its integration with other materials that can be fabricated via UHV fabrication processes—reactive metals such as Cobalt, for example —thereby paving the way for the fabrication of a broad range of organic-inorganic systems. Moreover, as the polarization reversal does not involve any energy-expensive molecular rotation mechanisms, CA is expected to have a reasonably small coercive field value. Thus, CA naturally becomes a desirable candidate for organic ferroelectric-based device applications, which inherently benefits from the advantages offered by organic materials along with high spontaneous polarization and small coercive fields.

Following the discovery of ferroelectricity in the crystal form of CA, several groups have started working on various aspects of the powder,[48–50] bulk single crystal,[51–59] self-assembled co-crystal[60,61] composite film[62] and thin film[63–70] forms of the molecule. Most of the works have focused on ferroelectric and optical properties of the macroscopic crystals of CA. While the realization of the potential of CA in device applications relies on the detailed studies of growth and ferroelectric properties in its thin film forms, such studies are just beginning to gain attention. In view of CA as a novel organic ferroelectric, and of its potential for future applications, we present an up-to-date concise survey covering most of the scientific activities that have been carried out on the single crystal and thin film forms of the material. To provide a general perspective to the field, we begin this review with a brief introduction to different types of organic ferroelectric materials and then focus on CA. Thereafter, the molecular structure of CA is discussed followed by the description of molecular organization in the crystal lattice. The subsequent coverage is divided into two parts. In the first part, we illustrate the origin and nature of ferroelectric polarization in the solid-state macroscopic single-crystalline form of CA and present a summary of ferroelectric characterizations performed so far on such single-crystals. In the second part, we discuss the studies carried out on thin films, including ultra-thin and relatively thicker polycrystalline films, followed by presentation of studies performed toward applications of CA films in nanoelectronic devices. Finally, we conclude the review with some perspectives on the future directions of explorations with this novel organic ferroelectric.

## 2. Types of organic ferroelectric systems
2.1. Polymer-based organic ferroelectrics

Ferroelectricity has been shown to exist in many polymeric materials such as polyureas, polyurethanes, nylons, poly-vinylidene fluoride (PVDF), and co-polymers of PVDF.[71–75] Of all the polymer-based ferroelectric materials, PVDF and its copolymer PVDF-trifluoroethylene (PVDF-TrFE) have received the maximum attention mainly due to their significant remanent polarization, short switching time, and thermal stability. Ferroelectricity in PVDF originates from the individual polar vinylidene fluoride molecules which carry the electronegative fluorine and electropositive protons at opposite ends of it. The application of an electric field results in the rotation of the molecules, which results in the reversal of the molecular dipole moment and hence the net polarization reversal in the molecular system. In the co-polymer PVDF-TrFE, the presence of trifluoroethylene group helps in the molecular rotation. In a particular growth phase, when the hydrogen and fluorine atoms are positioned opposite to each other, the remanent polarization is approximately 7 $\mu C/cm^2$ with a coercive field of around 90 V/m.[76–78] PVDF-based ferroelectric films have been very useful in the commercial sector as a ferroelectric material due to their abrasion resistance, durability, non-flammability, and radiation tolerance.[79–81] In the field of nanoelectronic devices as well, PVDF-based ferroelectrics have received a substantial attention.[7,24,82–84]

## 2.2. Charge transfer complexes

If a multi-component molecular material is composed of one electron-rich donor-type (D) unit and another electron-poor acceptor-type (A) unit, and neutral pairs of these two types are stacked together, an intermolecular/inter-unit charge transfer can occur within the donor-acceptor (D-A) system. The charge transfer can break the symmetry of an otherwise regular donor-acceptor type neutral arrangement to form a polar chain arrangement, with a polarity that is mediated by electrostatic interactions. The two energetically degenerate polar arrangements, $A^-D^+A^-D^+$ and $D^+A^-D^+A^-$ can represent the two states of ferroelectricity. Clearly, this polar state will occur only if it is energetically favorable. This reflects the energy cost in creating charged species, and the energy reduction due to the consequent attractive electrostatic interactions between them.[31,44,85]

Charge transfer complexes such as tetrathiafulvalene (TTF) –quinone (Qui) type systems are well studied.[32,86,87] Tetrathiafulvalene–p–chloranil (TTF–ChA) complex, where TTF serves as the electron donor and ChA as an acceptor, is a typical example of such a system, where the non-polar centrosymmetric neutral structure undergoes a neutral to ionic phase transition to create a polar ionic structure below 81 K.[88–90] This polar structure is ferroelectric with a remanent polarization of ~6.3 $\mu C/cm^2$ and a coercive field of ~5.4 kV/cm at 51 K.[33] The ferroelectric first-order phase transition is accompanied by sharp change in the dielectric constant.

## 2.3. Hydrogen bonded systems

If a donor-acceptor hydrogen bonded bi-component system can crystallize into a linear chain of alternating donor-acceptor pairs, it is possible that the dipoles in the acid-base pairs can be ordered and ferroelectricity may emerge. There are two possible ways in which ferroelectric ordering can be generated in the hydrogen bonded chains of a bi-component donor-acceptor system. (i) Both the donor and acceptor molecular units in the chain are individually neutral and ferroelectricity arises from the displacive behavior of H atoms in a charge neutral chain. (ii) An actual proton ($H^+$) transfer occurs between the donor and acceptor molecular units, giving rise to an ionic hydrogen bonded chain system in which ferroelectricity results from the ordering of dipoles created by a non-centrosymmetric arrangement of transferred protons.[44]

A good example of type (i) ferroelectricity is the supramolecular chain assembly of two π-conjugated molecular systems: phenazine (Phz) as the H-acceptor (base component) and chloranilic acid ($H_2ca$) as the H-donor (acid component). $H_2ca$ possesses two proton-donating O-H groups and Phz possesses proton-accepting N atoms on both side of the molecule. Thus, these two components co-crystallize to form alternating hydrogen-bonded acid-base pairs. At room temperature, the H atoms are positioned centro-symmetrically and the charge distribution in the entire chain is neutral. Below the critical temperature, the H atoms of the $H_2ca$ molecules become displaced toward the neighbouring Phz molecules and the Phz molecules get displaced toward the neighbouring $H_2ca$ molecules. The resulting inversion symmetry breaking gives rise to a net polarization of the supramolecular chain. The spontaneous polarization of the Phz-$H_2ca$ system is ~1.2 μC/cm$^2$ at 160 K.[34,35,39,91]. It must be noted that there is no intermolecular transfer of proton in such a supramolecular system.

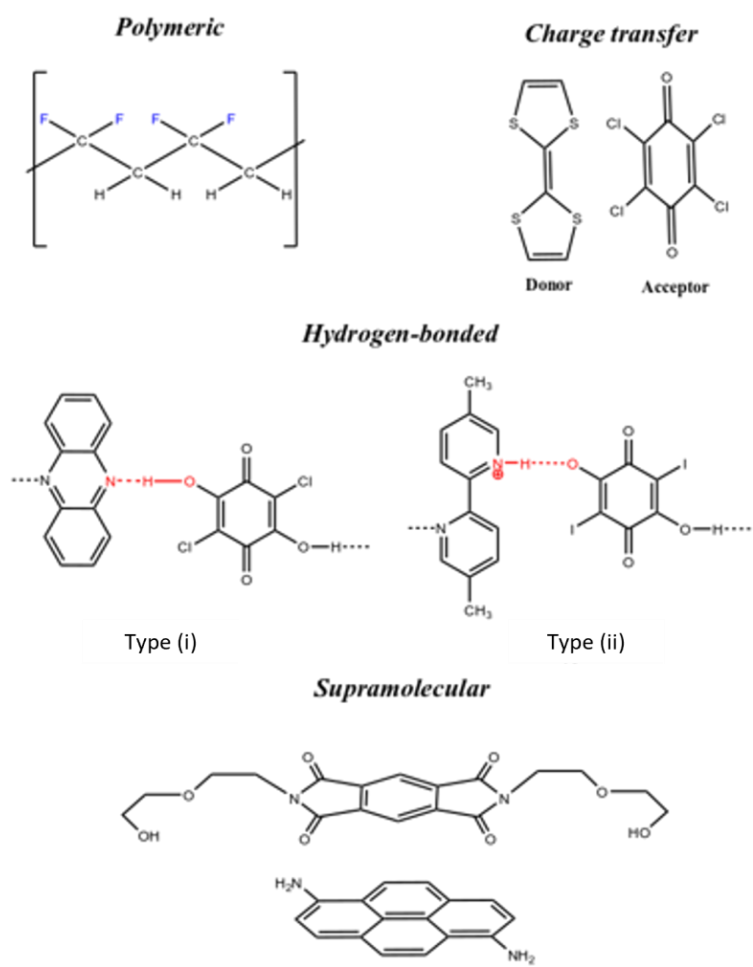

**Figure 2.** Representative examples of organic ferroelectric systems. Top row: PVDF (left), TTF-CA (right). Middle row: Phz-$H_2ca$ (left), [5,5-DMBP][$H_2ia$] (right). Bottom row: PMI derivative and TTF derivative. Atoms and bonds highlighted in red for the compounds in the middle row represent the intermolecular hydrogen bonds. Type (i) and type (ii) represent the two types of hydrogen bonded ferroelectric systems without and with the intermolecular proton transfer process, respectively.

Type (ii) on the other hand represents systems where, unlike in the case of type (i), the H-atom/proton of the hydrogen bond gets transferred from one molecular unit to another neighboring one during the ferroelectric reversal process. This transfer can take place in a concerted way; that is, without changing the chemical structure of any components. A nice example of such a system is the supramolecular assembly iodanilic acid ($H_2ia$) and the 5,5´-dimethyl-2,2´-bipyridine (5,5-DMBP) base. In the ferroelectric state of this system, the O-H bonds of $H_2ia$ and the N-H$^+$ bonds of 55DMBP are aligned along the same direction. Ferroelectricity results from the alternative polar ordering of O-H---N and O$^-$---H-N$^+$ bonds. During the polarization reversal, the H-atom undergoes intermolecular transfer between the N-atom of 55DMBP and the O atom of $H_2ia$. For all the H-atoms to undergo a concerted transfer, the π-electron system of both the base and acid units should also participate in the transfer process.[36,40] Interestingly, hydrogen-bonded ferroelectric systems need not necessarily be of multi-component nature. Single-component systems can also show ferroelectricity. For example, 2-phenylmalondialdehyde (PhMDA), 3-hydroxyphenalenone (3-HPLN), and cyclobutene-1,2-dicarboxylic acid (CBDC) and 2-methylbenzimidazole (MBI) are proton transfer-based (type ii) single-component hydrogen bonded ferroelectric systems.[37,38,43] However, the most remarkable example of type ii hydrogen-bonded system is CA, which was shown to be ferroelectric only rather recently.[42] What makes this particular system interesting is its high value of above room-temperature polarization (30 μC/cm$^2$), which rivals that offered by inorganic ferroelectrics.[43] Clearly, CA has the highest room-temperature polarization among organic ferroelectric materials. The large value of polarization makes this organic ferroelectric a prominent candidate for possible applications. Studies elucidating the origin and nature of ferroelectricity in CA and the possible applications of the material are presented in sections 3.2 and 3.3, respectively.

## 2.4. Supramolecular systems

The supramolecular design of molecular system deals with forming ordered molecular assemblies using non-covalent interactions. Some of the types of organic ferroelectric materials discussed above can be considered as supramolecular systems. However, the quest toward newer, robust, above-room-temperature ferroelectric materials have led ways to some advanced supramolecular systems where different mechanisms, such as charge transfer and hydrogen bonding are synergistically combined to design advanced organic ferroelectric materials with the help of chemical functionalization.[92] For example, Tayi et al.[41] have designed a room-temperature supramolecular ferroelectric charge transfer complexes based on pyromellitic diimide (PMI)-based electron acceptor and TTF based donor units. This strategy encompasses four different types of non-covalent bonding interactions— namely, hydrogen bonding, charge transfer, π-π stacking and van der Waals type interactions.

## 3. Croconic Acid
## 3.1. Structure of an individual Croconic Acid molecule

Croconic acid is an oxo-carbon ($C_nO_n^{2-}$)-based acid with five carbon atoms (n=5) (figure 3). It is a simple organic molecule composed of carbon, hydrogen, and oxygen atoms. Two hydroxyl and three ketone groups linked to the five carbon atoms of a cyclopentene skeleton form the structure of the molecule.

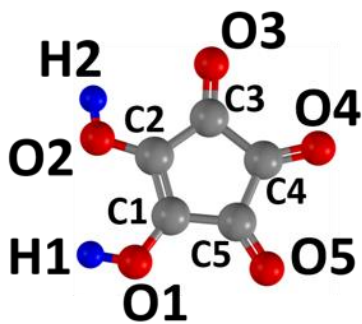

**Figure 3.** A CA molecule with all the atoms labelled is shown for reference.

A single croconic acid molecule is electrically polar with a dipole moment of 7-7.5 D (1 D = 3.336 × $10^{-30}$ C·m).[93],[94] The dipole moment is due to the electronegativity difference between the two hydrogen and oxygen atoms present in the hydroxyl groups attached to the C=C unit. Furthermore, the relative orientations of the H atoms in the hydroxyl group define the isomeric forms with distinct dipole moments. For example, both H atoms symmetrically pointing outward or inward gives rise to an achiral isomer. On the other hand, a chiral isomer (figure 3) develops when one of the H atoms points inwards (atom H1) while the other points outwards (atom H2).

The total density of states (DOS) (panel 1) and partial-DOS (PDOS) of constituent atoms (panel 2 for H, panel 3 for C and panel 4 for O) obtained from Density Functional Theory (DFT) calculations are shown in figure 2 of reference 99. The strong overlap between the PDOS peaks in the valence band corresponding to 'p' orbitals of C and O atoms, shown in the third and fourth panels respectively, is indicative of an electron delocalization or a covalent bonding between the C and O atoms. Furthermore, the absence of any PDOS corresponding to the 's' orbitals of H atoms at the bottom of conduction band underscores the ionic nature of the O-H bonds.

### 3.2. Solid state bulk single crystal form of Croconic Acid

The slow evaporation of the aqueous solution of croconic acid in 1N hydrochloric acid results in the crystallization of semi-transparent, elongated slab-like yellow crystals with several 100s of micrometer in length and several 10s of micrometers in thickness. Although croconic acid is known to exist for a long time, its crystallization was successfully carried out only very recently by Braga et al[95].

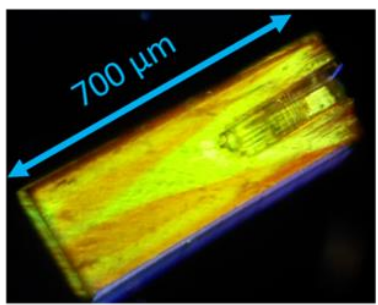

**Figure 4.** Optical images of 1N solution grown crystals of croconic acid.

3.2.1. Molecular structure in the crystal form

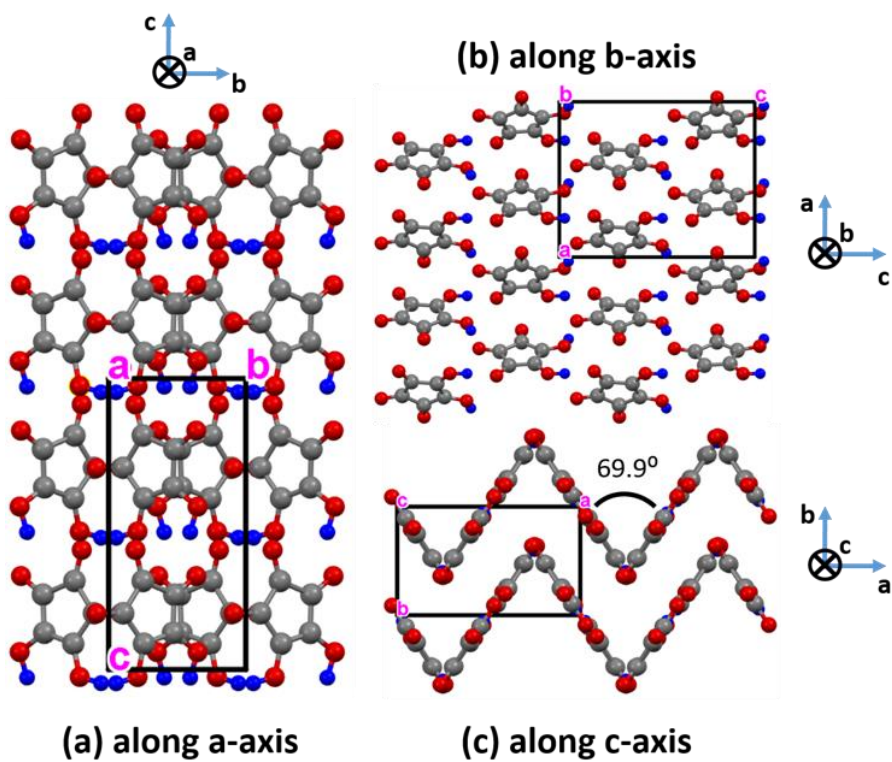

**Figure 5.** Views of CA crystal structure with two stacked layers along three orthogonal axes. The axes mentioned as a cross symbols represent directions into the plane of paper. Black boxes represent unit cells. Red, blue, and grey balls represent oxygen, hydrogen and carbon atoms, respectively. The structures are generated using the Pca2$_1$ structure (identifier number=753043) of the Cambridge structural database.

Croconic acid crystalizes in the *Pca2₁* space group belonging to Orthorhombic crystal system. Each unit cell is composed of four molecules of CA. In the solid-state crystal form, croconic acid molecules form sheets of hydrogen-bonded networks. Each sheet consists of zigzag planes (figure 5c, 6a) with an angle of 69.9º between the adjacent planes. These zigzag sheets are stacked on top of each other to form the macroscopic crystal and are shifted in order to avoid an overlap of molecules across the sheets. Within each plane in a sheet, CA molecules are again arranged in a zigzag manner (figure 5b, 6a), forming a linear chain of hydrogen bonds along the edge of adjacent planes in the zigzag sheet and parallel to the crystallographic c-axis (figure 6c). The formation of these hydrogen bonds is facilitated by the presence of oppositely positioned hydrogen bond donor (H1-O1 and H2-O2 groups in figure 3) and hydrogen bond acceptor (C4=O4 and C5=O5 groups in figure 3) groups within the network of CA molecules formed across the zigzag planes in a sheet. Each CA molecule has a chiral form (figure 3) and is hydrogen bonded to four neighboring molecules, two in its own plane and the remaining two in the adjacent plane forming a tetrameric ring as the unit cell (rectangular boxes in figure 5b represent the unit cells). The structural information within the crystals has been obtained from detailed structural characterizations including X-ray diffraction, neutron scattering, and theoretical studies on CA crystals.[49–51,53,66,96,97] The views of the crystal structure along the main crystallographic axes are presented in figure 5 and along arbitrary directions in figure 6 for a clearer understanding of the molecular arrangements in CA crystals.

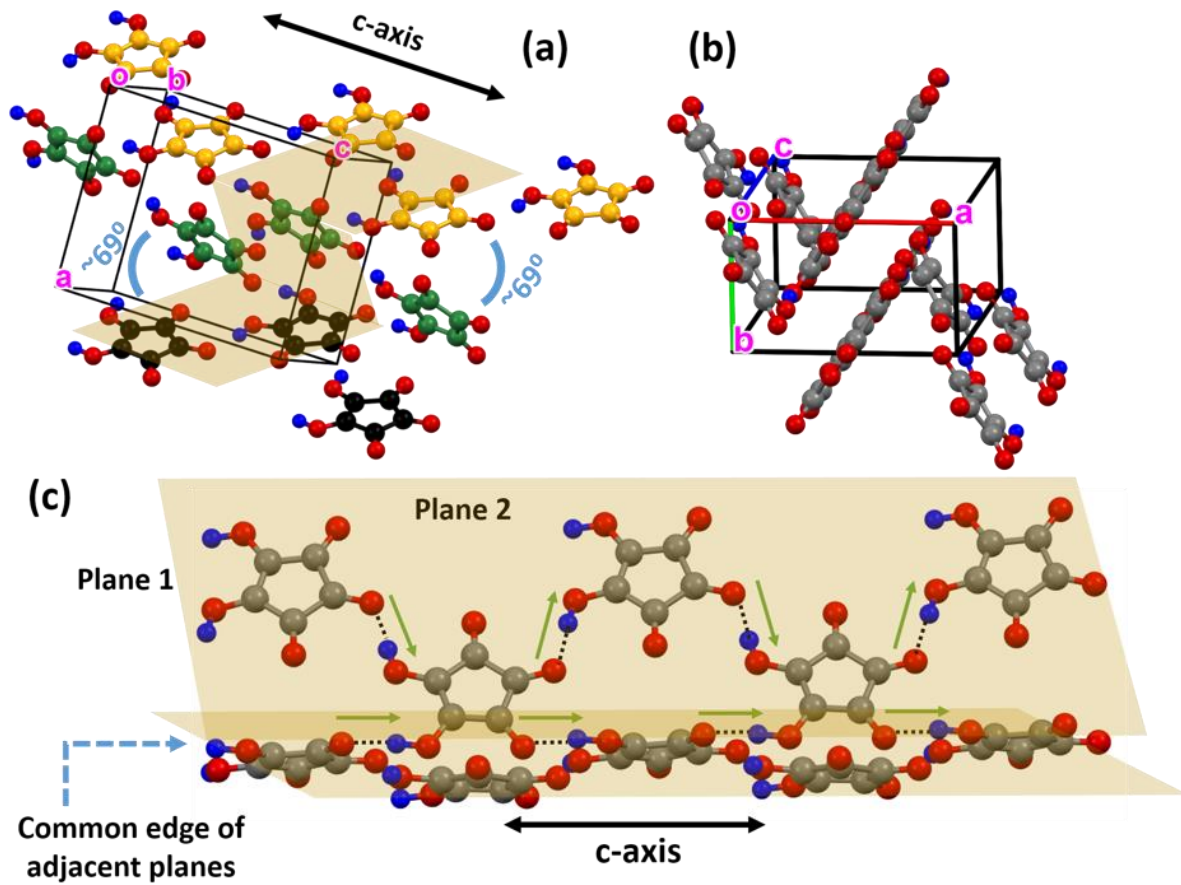

**Figure 6.** Zigzag sheets of a molecular networks viewed along arbitrary directions. (a) shows 3 consecutive planes in one sheet with different number of molecules in each plane. For ease of distinction, the 3 molecules in the lowest plane are shown with black carbon atoms. Similarly, the 4 and the 5 molecules of the middle and the topmost planes are shown with green and yellow carbon atoms, respectively. An angle of 69⁰ between the consecutive planes is indicated. The skewed rectangles with yellow shade represent the three consecutive planes forming the zigzag sheet. (b) shows two consecutive sheets with two planes in each sheet and 3 molecules in each plane. 'o' stands for the origin of the coordinate system. oa, ob, and oc indicate the direction of a, b, and c axes, respectively. (c) shows two adjacent planes with 5 molecules in each plane. The dashed lines indicate the intermolecular hydrogen bonds, and the green arrows show the dipole moments arising out of the asymmetry in the O-H---O bonds. All the dipole moments along the common edge of two planes point along the same direction, thereby adding up. The direction of net polarization of this network configuration (towards right hand side) is opposite to the relative positioning of hydrogen atoms with respect to the individual host molecules (towards the left-hand side). The structures are generated using the Pca2₁ space group structure (identifier number=753043) from the Cambridge structural database.

### 3.2.2. Origin of the ferroelectric order in croconic acid

Although a CA molecule is itself polar owing to the presence of electropositive H atoms, the polar nature of bulk CA crystals has a more intriguing origin with contributions from both ionic and electronic characteristics of the hydrogen-bonded molecular networks. The ionic characteristic of the polar ordering in the crystal is due to the non-centrosymmetric positioning of the H atoms of the molecules with respect to the hypothetical centrosymmetric paraelectric structure (figure 7b). As shown in figures 6c and 7a, out of the four intermolecular hydrogen bonds that each molecule participates in, two lie along the common edge of the adjacent planes in the zigzag sheet and are parallel to the c-axis. The asymmetry in the positions of the H atoms in these hydrogen bonds (O---H-O) is additive along the c-axis, thus it is expected that the major component of the net polarization due to the non-centrosymmetric network structure is also along the c-axis. The unidirectional dipole moments arising due to the asymmetric intermolecular hydrogen bonds are shown as green arrows in figure 6c.

Detailed first-principles calculations based on the non-centrosymmetric positioning of hydrogen atoms suggest that the ferroelectric structure is energetically more stable than the paraelectric one, resulting in a polarization of 26 $\mu C/cm^2$ directed along the c-axis which is more than the sum of all the individual molecular dipole moments (~21 $\mu C/cm^2$)[42]. The calculated polarization value can go up to ~32 $\mu C/cm^2$ depending on the exchange correlation function used in the calculation.[98] The fact that the net polarization of the network is more than the sum of molecular dipole moments is indicative of the contribution of intermolecular interactions to the net polarization of the structure. Indeed, the charge-density difference between the actual network of the crystal structure and that of individual molecules with all the atoms at respective positions reveal substantial intermolecular charge rearrangements that take place in the crystal network.[42] The Bader charges on each of the atoms of a molecule in a crystalline CA network (table 1 in reference 99) obtained from first-principles calculations are also suggestive of the partial positive charge on the electron-releasing and partial negative charge on the electron-attracting groups.[99]

On the other hand, the electronic contribution to the polar ordering is expected to have its origin in the non-centrosymmetric positioning of the π electron unit and the associated electronic charge density in the CA molecules in a crystal. Theoretical calculations suggest that a significant contribution (~80%) to the polarization of the crystalline CA structure is of electronic origin.[43]

An electric field can induce polarization modulations in a CA crystal that can be attributed to the protonic and the electronic mechanisms. On the other hand, an ultrafast (timescale of ps) electric field may have different influence on the two mechanisms, as the domain wall motions, which can cause the polarization modulation due to the ionic/protonic motions, occur on a much slower time scale (ms) as compared to the ultra-fast scale (ps). This idea can be employed to utilize ultrafast experimental techniques to probe the exact contributions of electronic and protonic mechanisms. For example, an ultrafast Second Harmonic Generation (SHG) probe can be employed to detect polarization modulations and optical reflectivity probe can signify modulations in π-electronic densities in the molecules.

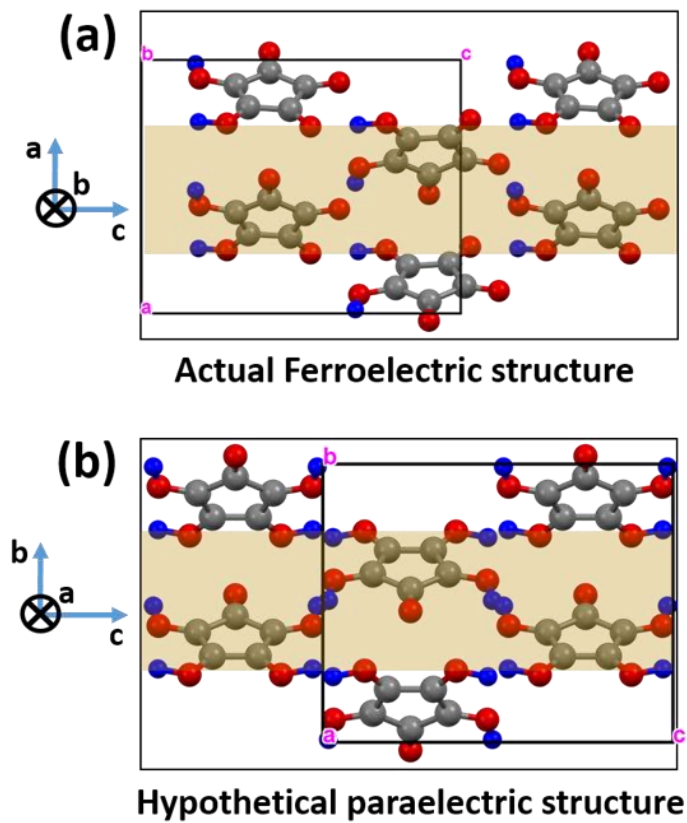

**Figure 7.** Structural comparison between the non-centrosymmetric actual ferroelectric crystal structure (a) and the hypothetical centrosymmetric paraelectric structure (b). The shaded yellow regions indicate one plane of the zigzag sheets with a two-dimensional view. Note the difference in the directions of axes for both cases. The structures are generated using the Pca2$_1$ space group structure (identifier number=753043) for ferroelectric structure and Pbcm space group structure (identifier number=753044) from the Cambridge structural database. Rectangular boxes represent unit cells.

The nature of electronic origin of the ferroelectric order has been studied by Sawada et al.[54] using ultrafast optical reflectivity, and SHG spectroscopy in conjunction with DFT calculations. The ultrafast electric field-induced electron density modulation causes the photo-excitation of the electronic ground states and the corresponding electronic transition (lowest π - π* transition) experimentally gives rise to a peak in the optical reflectivity plot at a photon energy of ~3.2 eV.[54] This peak in the optical reflectivity also signifies the modulations in the electronic densities within the molecules around the particular photon energy.

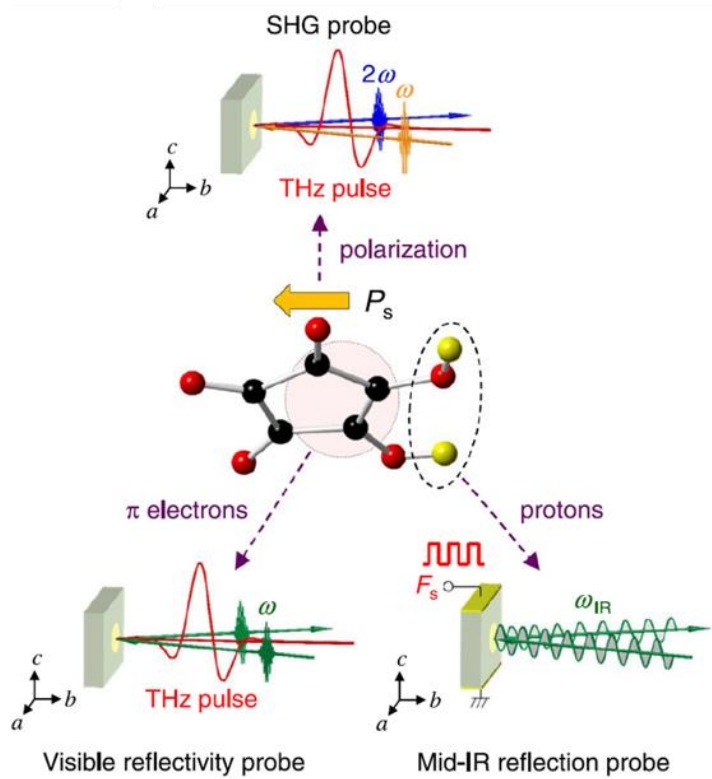

**Figure 8**. Schematics of the experimental setup for a combined ultra-fast IR-SHG-VIS spectroscopy of a CA crystal. Figure is reproduced with permission from reference 57, CC BY 4.0 (http://creativecommons.org/licenses/by/4.0/).

Further, DFT calculations show that the dipole moment of an individual CA molecule in the lowest photo-excited state is opposite to that in its ground state. This results in a large second-order optical non-linearity in CA crystals, which has been experimentally observed with the SHG spectroscopy.[54] The fact that the dipole moment of individual molecules can be tuned optically by photo-excitation of the energy states of the molecules in the crystal also strongly advocates for an electronic nature of the polarization in CA crystals.

Furthermore, the electronic nature of the polarization has been probed experimentally by Miyamoto et al.[57] by an interesting combination of multiple ultrafast probing techniques, IR+SHG+VIS (IR=Infrared, VIS=optical) spectroscopy (figure 8), while the crystal is subjected to ultrafast terahertz (sub-ps) electric field pulses. They observed a close correlation between the time-evolution of the SHG signal intensities and that of the applied tera-hertz electric field, which signifies the tera-hertz response of the polarization in CA. While it can be presumed that such a fast response should correspond to an electronic mechanism of ferroelectricity, the utilization of IR probe, which is employed to detect the possible protonic displacements involved in a polarization modulation process, validates the presumption by detecting negligible contribution of protonic displacement to the polarization modulation.

This thus suggests that the observed polarization modulation upon ultrafast electric field excitation is due not to the motion of H ions but to the modulation in the π-electron density of the molecular system. It indicates that in the steady state too, the net polarization of a crystal has significant contribution from the electronic nature of the molecular system.[57] Moreover, the fact that the polarization in CA can be modulated at ultrafast speeds may open up possible ultrafast applications of this organic ferroelectric.

### 3.2.3. Polarization reversal mechanism in croconic acid

In conventional organic ferroelectrics, electric field-induced polarization reversal occurs via bulk molecular rotation, as observed even for a croconic acid derivative, dimethyl croconate. In contrast, no molecular rotation takes place in CA. Rather, the polarization reversal takes place via a coherent proton tautomerization. Tautomerization is a common phenomenon in organic chemistry where energetically equivalent structural isomers of a chemical compound interconvert within each other. The most common of several tautomerization processes is keto-enol tautomerization (H-O-C=C and O=C-C-H). While molecular rotations may not always be energetically favorable due to steric hindrance in molecular systems, proton tautomerism on the other hand can be achieved/triggered at lower energies.

Within a network of hydrogen bonded molecules, the synchronized polarization reversal of the non-centrosymmetrically ordered intermolecular dipole moments occur through a co-operative keto-enol tautomerism. Each molecule simultaneously works as a donor and acceptor of pairs of protons with the two electron-releasing (hydroxyl) and two electron-attracting (ketone) groups working as acceptor and donor of protons, respectively. The two keto groups tautomerize to two enol groups and vice versa. In this way, the protons of the entire network co-operatively move and the net polarization of the crystal is reversed along the c-axis. This process is facilitated by a proper relocation of the delocalized π-bond in the central oxo-carbon system while the 5$^{th}$ keto group of each molecule stays neutral without participating in the intermolecular H-transfer.

Such a coherent tautomeric proton transfer in a molecular network is schematically shown in figure 9, which shows a portion of the hydrogen-bonded network in one sheet of the CA crystal. The central CA molecule is hydrogen-bonded to four neighboring molecules. The network shown on the left (figure 9a) has a net polarization oriented toward the left-hand side that, after polarization reversal, flips 180⁰ turn to align toward the right-hand side (figure 9b). The pairs of hydrogen ions/protons of the enol groups (as indicated by the red stars in figure 9a) from two neighboring molecules on the left of the central molecule get transferred to the central molecule in a synchronized way, tautomerizing two of its keto groups into their enol forms (red stars on figure 9b). Similarly, the two hydrogen ions/protons in the enol groups of the central molecules get transferred to two different molecules on the right-hand side. This tautomerizes their keto groups into enol forms, while the enol groups get stabilized in their keto form with the help of the delocalizing π-electron center. It must be noted that these processes are not sequential, but instead are synchronized throughout the network. This is how each molecule in the network participates in a collaborative process of synchronized proton transfer to reverse the polarization of the entire network.

Since a significant contribution to the steady state polarization of the CA network is of electronic origin, this electronic contribution must also reverse its directionality in synchronization along with the proton transfer upon the net polarization reversal in the crystal. This occurs in conjunction with the tautomeric proton transfer process by virtue of transfer of the π-bond and the associated electronic charge density of the central ring in each molecule to the opposite side of that ring.

The electronic density in each molecule in a hydrogen bonded network of CA is coupled to that of the neighboring molecules via resonance-assisted hydrogen bonds.[100] Thus, an interesting question may arise at this point regarding the electronic contribution during a polarization reversal. Is it possible to optically trigger the electronic density and achieve a stable and synchronized protonic displacement across the molecules in a macroscopic region of a CA crystal, thereby reversing the net crystal polarization by optical means? Indeed, Iwano et al.[56] have tried to investigate this aspect by ultrafast optical pulse-pump and SHG-probe-based measurements in conjunction with theoretical calculations. They experimentally observed that an ultrafast optical signal is able to modulate the polarization (decrease of polarization due to modulation in π-electronic density) and could theoretically estimate the loss of polarization. Further, their calculations suggest the existence of possible energy pathways via which ultrafast optical π-π* electronic excitations in individual CA molecules can trigger the displacement of the hydrogen ions to nearby molecules in a molecular network along the c-axis and result in the polarization reversal of a macroscopic region.

Along with the ionic and electronic mechanisms responsible for the existence and reversal of steady-state polarization, it is also important to recognize the co-operative and additive relation between the above two mechanisms, which essentially stabilizes large polarization values in CA. In the steady state, the π-electron density in each molecule is asymmetrized in a direction opposite to that of the protonic/ionic asymmetry due to the intermolecular hydrogen bonds associated with the particular molecule. Thus, the dipole moment generated due to the electronic asymmetry gets added to the protonic one and results in an enhanced net polarization. Similarly, during the polarization reversal, the electronic displacement occurs in a direction that is opposite to that of the protonic/ionic one. This aids the polarization reversal process by stabilizing the net reversed polarization. It is clear that the addition of the electronic and the protonic contributions is most effective when the relative angle between the directions/axes of the two asymmetries is exactly 180⁰. Any inclination between these two directions will reduce the electronic contribution to the net polarization, and the minimum addition occurs in the case of a parallel orientation between the two asymmetry axes. Indeed, there are examples of organic ferroelectric molecules, such as 3-anilinoacrolein anil, where the direction of electronic asymmetry is inclined at an angle (90⁰ in this particular case) to the protonic displacement direction. In such molecules, the electronic and protonic contributions to the net polarization do not add up, hence resulting in reduced polarization values.[43]

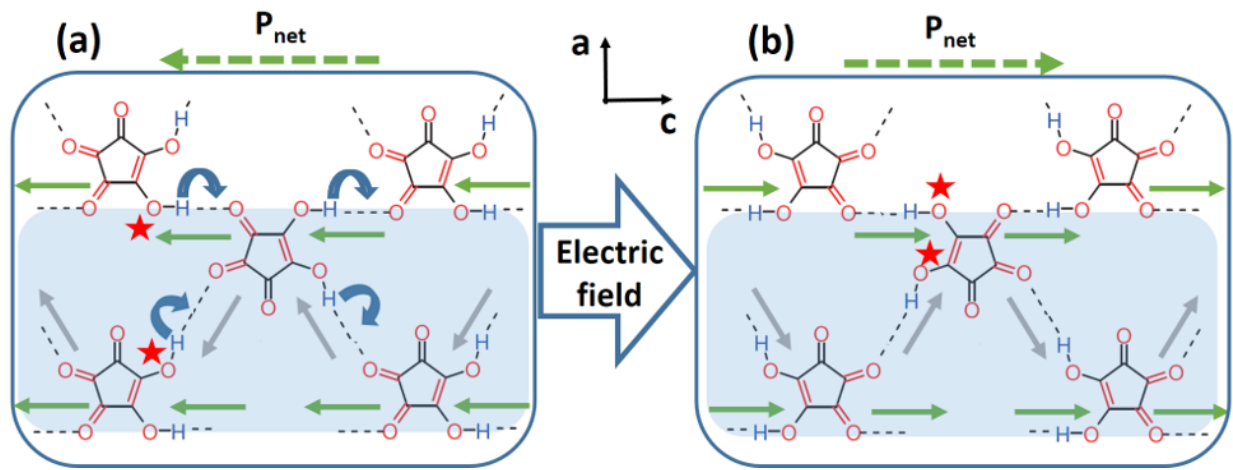

**Figure 9**. Schematics of proton tautomerism based-polarization reversal process in a hydrogen bonded network of CA molecules. (a) and (b) show the molecular network in opposite states of polarization orientation along the c-axis. Thick curved blue arrows on the H atoms surrounding the central molecule represent the direction of tautomeric motion of the protons during the process of reversal. The green shades represent one plane along c-axis in the particular zigzag sheet of CA. The red stars indicate two of the four enol groups that are involved in the tautomerization of the central CA molecule. Green and grey arrows indicate the intramolecular dipole moments along the c-axis and some arbitrary axes, respectively.

Lastly, the moderate strength of the hydrogen bond in the CA structure also plays a crucial role in allowing the polarization reversal to occur via proton transfer. Hydrogen bonds that are too weak would not be able to sustain the non-centrosymmetric positioning of the protons/hydrogen atoms, bonds that are too strong will not allow for the proton transfer to occur, thereby hindering the polarization reversal process.

### 3.2.4. Ferroelectric properties

Polarization hysteresis loops obtained by Horiuchi et al.[42] by the application of electric fields with triangular-bipolar waveforms (pulse amplitude modulated) along the c-axis of a CA crystal are shown in figure 10a. The remanent polarization value reaches a maximum of around 30 $\mu C/cm^2$. These CA crystals were optimized by Horiuchi et al.[43] by repeated poling and thermal treatments before the hysteresis measurements.

Usually, during the growth of crystals from a solution, multi-domain structures form with ferroelectric domain walls separating the domains. The polarization charges on domain walls create huge depolarizing fields, which are compensated by the embedding of dense mobile charges and charged mobile or immobile defects in the crystals during growth at room-temperature. These pinned defects and charged domain walls result in an imperfect and incomplete switching of bulk polarization due to a relatively slower motion of the charged domain walls. This results in the suppression of the observed remanent polarization values.[101] However, by thermal annealing and repeated poling procedure, these defects can be removed, the remanent polarization increases and a faster switching is obtained. In this way, Horiuchi et al.[43] have increased the remanent polarization of CA crystals from 20 µC/cm² to ~30 µC/cm², increased the switching frequency without significant loss of polarization, reduced the frequency dependence of the remanent polarization, and improved the squareness of the polarization hysteresis loops without significantly increasing the coercive field (figures 10b-d).

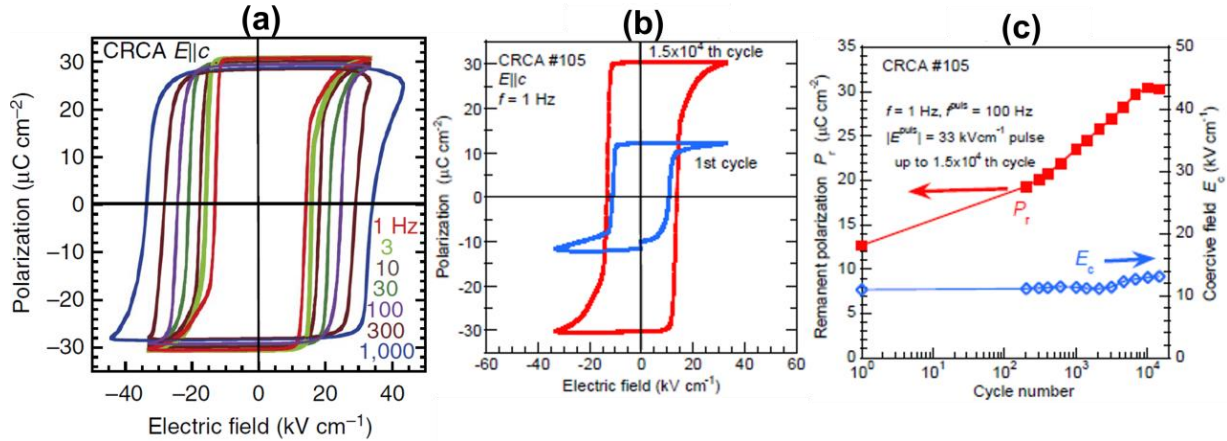

**Figure 10.** Standard ferroelectric characterizations of CA crystal. (a) Ferroelectric polarization hysteresis loops measured on different CA crystals using a triangular AC electric field after proper optimization through thermal and electrical treatments, respectively. Remanent polarization reaches up to 30 µC/cm² and its reduced frequency dependence is evident. (b) Polarization hysteresis of another similar crystal before and after repeated polarization reversal optimization procedure. (c) shows how the remanent polarization and coercive field values evolve with repeated cycles of polarization reversals. Figures are adapted with permission from reference 44 (CC BY 4.0 http://creativecommons.org/licenses/by/4.0/).

The effect of poling a crystal was optically investigated by Horiuchi et al.[42] SHG imaging in the absence of electric field was carried out both in the pristine and poled state of a CA crystal. An increase in SHG intensity after poling the crystal along c-axis was observed, signifying ferroelectric activity. Randomly distributed anti-parallel ferroelectric domains in the pristine state resulted in the destructive interference owing to the anti-phase SHG light near the domain walls. In the poled state when the domains aligned in a parallel fashion, such a destructive interference was subdued and thus the increase in SHG intensity was observed. Further, the modulation of the SHG intensity from CA crystals have also been quantitatively measured by others both in the reflection and the transmission modes with femto-second laser sources.[57],[54]

The multi-domain structures of the as-grown crystals of CA and the domain propagation under applied electric field have been imaged with terahertz (THz) radiation imaging by Sotome et al.[55] Femto-second laser pulses in the THz frequency range impinge onto the crystal and the radiated

THz electric fields, which are emitted parallel to the crystal c-axis owing to the presence of a non-centrosymmetric order in the crystals, are measured in transmission mode. The magnitude and direction of the crystal polarization is locally mapped by respectively measuring the amplitude and phase of the emitted signal. The 180⁰ phase reversal of the THz electric field upon polarization reversal of the crystal, obtained by applying external electric field along the c-axis, demonstrates the coupling between the polarization and the emitted THz electric field. The spatial map of the radiated THz field phase reveals the ferroelectric domain structure of the crystals.

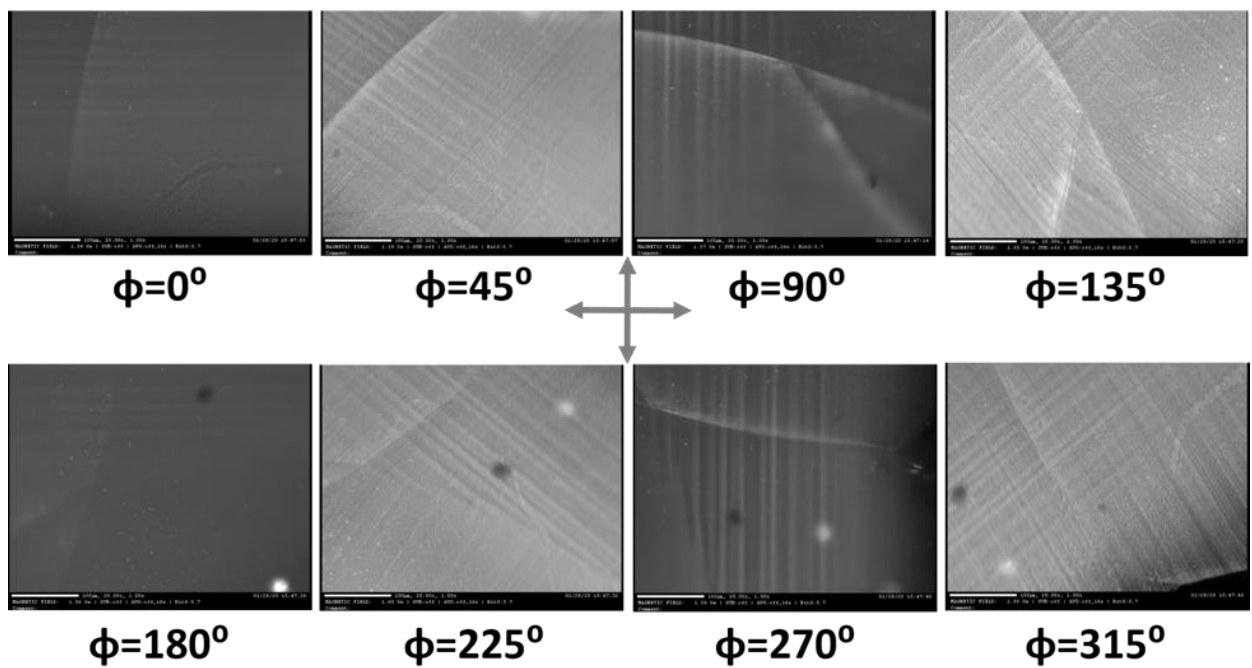

**Figure 11.** Optical polarimetry contrast under crossed analyzer-polarizer configuration. The angle ϕ denotes the angle of sample rotation with respect to the orientation of the polarized light. The scale bar represents a length of 100 µm. Figure is reproduced from reference 102.

The domain wall propagation dynamics is captured by THz mapping of the crystal while increasing the external electric field strength. It is observed that, even for a pristine crystal with a domain wall perpendicular to the c-axis, domain wall propagation does not occur along the c-axis, but rather as a pair of domain walls along the a-axis. This is indicative of the strong anisotropy of the hydrogen-bonded dipolar arrangement in CA crystals along the c-axis, which makes it difficult for the cooperative proton and π-bond switching to occur along that axis during domain wall propagation.

Another important direction in the studies on the ferroelectric domains in croconic acid crystals involve the optical polarimetry measurements carried out by Mohapatra et al,[102] where stripe and knitted domain patterns have been observed on macroscopic crystals under polarized light (figure 11). Such patterns appear to be electrically active, as indicated by the change in the relative fringe pattern contrast upon rotation of the sample under a crossed analyzer-polarizer configuration by using a polarized light. Although an understanding of the exact nature of such domains requires further investigations, via combined electrical and optical means, the presence of these domains patterns on CA crystals indicate the interesting possibility of such crystals harboring exotic domain structures.

### 3.3. Thin films of croconic acid

One of the practical advantages that CA offers is its vacuum sublimability, which permits the fabrication of thin films under high or ultra-high vacuum and the integration of the process with state-of-the-art ultra-high vacuum thin films fabrication systems. While thin films of almost all other organic ferroelectrics can be grown only by using ex-situ atmospheric techniques such as drop casting or spin coating methods, the possibility to deposit CA thin films under ultra-high vacuum conditions makes CA a suitable material for high-quality device fabrication in large-scale commercial applications. In particular, fabrication of devices that integrate reactive materials, such as ferromagnetic metals, would be possible without degradation due to glove-box or atmospheric contamination.

Several groups have exploited this advantage to grow and study the properties of CA thin films. These studies, which have only recently begun, can be categorized mainly into two sections: *(i)* studies on ultra-thin films and *(ii)* studies on polycrystalline thin films. The former type of studies deals with the growth and structural properties of a single molecular layer of CA, which are commonly probed by Scanning Tunneling Microscopy (STM) and supported by DFT calculations. The later type of studies involves the fabrication of polycrystalline films with thicknesses in the range of several 10s of nanometers, to optimize the growth conditions and to study the ferroelectric properties of such films, toward possible device applications.

### 3.3.1. Ultra-thin films
#### 3.3.1.1. Molecular structure

When molecules are vacuum-grown onto metal substrates, the molecular crystallization process can strongly differ from crystallization in a liquid solution. The structure of individual molecules and the entire network or layer of molecules that can form on a substrate surface are determined by several factors, such as the energy of the evaporated molecules in the gas phase, and the surface chemical reactivity and temperature of the substrate surface. Furthermore, the molecular arrangement may depend on the substrate material surface, the thickness of the substrate surface (in the case of ultra-thin substrates), and the thickness of the deposited molecular layer. Thus, for CA molecules also it is expected that the molecular organization on substrate surfaces may be different from that of solution-grown bulk crystals. For example, for a thickness in the sub-monolayer regime, individual CA molecules may adsorb on the surface and upon increasing the thickness, several molecules may develop intermolecular hydrogen bonds to form dimers, trimers or extended molecular networks on the surface. Furthermore, since CA is a planar molecule, it may be expected that the planar structure is retained when the molecules are constrained to adsorb on a flat surface. However, the exact details of the molecular orientation depend on the particular type of monomer or molecular networks formed and the type of interaction that takes place at the substrate-molecule interface.

Depending on the relative positions of the two hydrogen atoms of a molecule, a total of 6 monomer conformations of CA are possible for a planar orientation (supplementary figures S2-S5 in reference 63). It is evident from figure S2 that, out of the six monomer conformations, only S2.b and S2.c are chiral and the rest including the minimum energy configuration S2.a are achiral. Here, the chirality of an individual molecule depends on the molecular conformation and is defined by the relative positions of H atoms in the individual isomers of the molecule. The asymmetric positioning of H atoms with respect to the central pentagon constitute the chiral conformation, while a symmetric positioning constitutes the achiral one. Various possible types of low-energy dimers, trimers and hexamers formed from the chiral and achiral monomer conformations, as optimized by first principles calculations to lie in one plane, are displayed in supplementary figures S3, S4 and S5, respectively in reference 63. It can be noted that, in the case of trimers some configurations tend to preserve the overall chirality of the entire trimer unit.

Experimentally, the growth of two-dimensional (2D) layers of chiral and achiral networks supported by metallic substrate surfaces have been observed, contrary to the densely packed zigzag sheets of molecular networks in the three-dimensional crystals. The substrate supported layers are shown in figure 12 for sub-monolayer growth on the surface of Au(111) and in figure 1 of reference 63 for Ag(111).[63] Figure 1a of reference 63 shows the isolated triangular clusters of chiral dimers for very low sub-monolayer coverage, and figure 1b shows the extended honeycomb network for one monolayer coverage (post annealing at 350 K) on Ag(111) surface. The observed dimer structures are the fundamental building blocks of all the observed porous network structures on Ag(111) surface. Similarly, on Au(111) surface, 2-dimensional sheets with Kagome lattice structures (figures 12 a, b), as opposed to the observed honeycomb structures on Ag(111) surface, have been observed.[64]

Close analyses of figures 1 a and b of reference 63 reveal that the molecules on a Ag(111) surface in the extended networks are in chiral form, corresponding to the dimer configuration S3.f depicted in figure supplementary figure S3. However, extended networks composed of achiral CA conformations are also observed, albeit with a very low frequency. On the other hand, the extended structures on Au(111) for one monolayer thickness are achiral (figures 12 a, b). These structures are composed of CA trimers that are fundamentally different from those found on Ag(111), and correspond to the trimer configuration S4.a shown in supplementary figure of reference 63. Thus, it should be noted that the basic monomer units which are the building blocks of the networks on Ag and on Au surfaces are different; it is chiral on Ag(111) (figure S2.b) and achiral on Au(111) (figure S2.a).

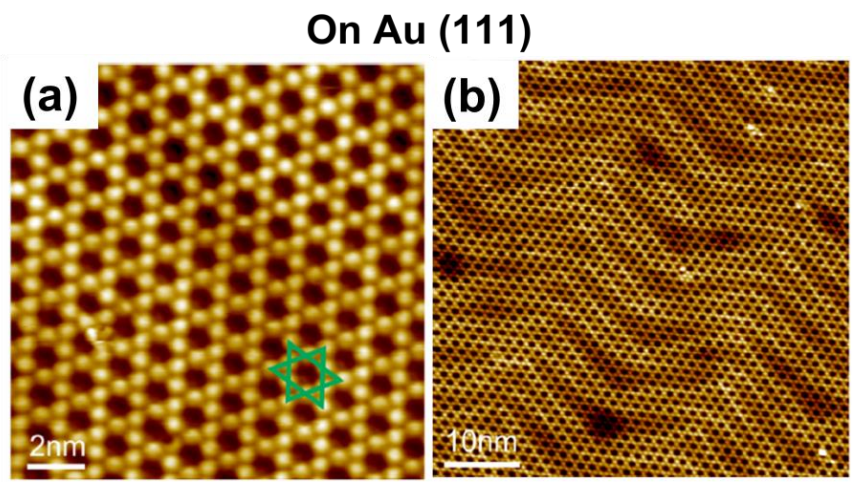

**Figure 12.** Substrate supported two-dimensional (2D) growth of CA networks as probed by Scanning Tunneling Microscopy (STM) imaging. (a) and (b) represent small and large scale surface morphology on Au(111) surface. Figures are adapted with permission from reference 64 (American Chemical Society, Copyright (2015)).

Experimental and computational analyses suggest that the observed structural difference is due to the weaker influence of the Au(111) surface on the molecules as compared to the intermolecular interaction strength. DFT calculations of modeled planar networks in vacuum (constrained to lie in a plane without any metal substrate) suggest that there is an energy difference of 0.2 eV between the two basic monomers, S2.a and S2.b (supplementary figures in reference 63), that constitute the networks on Au(111) and Ag(111) surfaces (figure 12), respectively, with the monomer on the Ag surface having higher energy. The stability of the lower energy monomer-based networks, reveals the weak influence of substrate surface on the network formation of CA molecules and is indicative of a stronger intermolecular interaction as compared to the substrate-molecule interaction on Au(111) surface. Similarly, the stabilization of higher energy monomer based on networks on Ag(111) surface indicates a stronger substrate influence and stronger substrate-molecule interaction strength.[64]

### 3.3.1.2. Ferroelectric Polarization and its reversal

An individual CA molecule is expected to have a dipole moment of approximately 7 D in the gas phase. However, this value depends on the particular monomer conformation of the molecule. As shown in supplementary figure S2 (reference 63), the two monomer conformations with the lowest energy (S2.a and S2.b) have significant calculated dipole moment values, although the second-lowest energy monomer (S2.b) has a larger dipole moment. It is interesting to note that S2.b is the same chiral monomer whose extended networks are dominantly stabilized on Ag(111) surface. Furthermore, the individual chiral dimer units on Ag(111) and achiral trimer units on Au(111) surface are non-polar because the dipole moments of the constituent monomers cancel each other out. The resulting entire networks formed from these dimer and trimer units are therefore non-polar. These stable non-polar structural arrangements have been found to form only in films grown on certain substrate surfaces. As the structure in the solution-grown crystals is different, the above structures could be exclusive to thin films.

Detailed simulations including molecular dynamics (MD) and DFT have been performed to explore whether polar two-dimensional networks on these metallic surfaces that can host proton transfer mechanism and give rise to a net polarization. Interestingly, Kunkel et al.[63] have observed, in molecular dynamics simulations of CA layers constrained to lie in a plane without any substrate surface, that the experimentally observed non-polar networks on Ag(111) surface can undergo spontaneous dynamical structural changes and stabilize a polar network after cooperative hydrogen atom transfer and rearrangement in the absence of any electric field. Such stabilization of polar networks has also been observed in the CA layer when the simulation is performed with two monolayers of Ag as a substrate surface in proximity to the CA layer.

However, these spontaneous cooperative hydrogen atom transfers are not necessarily concerted; rather it could be step-wise, as suggested by the detailed time-dependent MD simulations tracking of all the hydrogen bond lengths in a network similar to that found on Ag(111) surface.[103] In the presence of an in-plane electric field, however, the collective proton transfers may become partially concerted.

Similarly, for the non-polar networks observed on the Au(111) surface, calculations show that collective hydrogen transfer may result in a polar network with comparable energy, as the energy barrier between the two networks is comparable to that computed between the polar and non-polar configurations of CA crystals.[103] However, MD simulations concerning the energetics of the hydrogen transfer are yet to be performed.

No matter how interesting the feasibility of ferroelectricity in substrate supported ultra-thin films of CA may sound, the experimental verification of on-surface ferroelectric properties poses astounding challenges. The commonly used electromechanical method of characterizing local ferroelectric properties of thin films uses a local tip to generate a vertical electric field, so as to create a mechanical deformation of the local region in a thin film via the inverse piezoelectric effect. The mechanical deformation is then detected via photosensitive detection of the tip-hosting cantilever. However, this technique may not be effective for studying atomically thin systems, especially when the net polarization of the planar sheet type molecular network structures, as found on Au(111) and Ag(111) surfaces, lies in the film plane due to the planar orientation of dipole moments of the constituent monomers, dimers or trimers. It may thus be difficult to probe the ferroelectricity with scanning probe technique. For the same reason, it also becomes challenging to probe the ferroelectric reversal properties of the ultra-thin films using STM, which is otherwise a very effective nanoscale probing technique.

Nevertheless, both for a single CA molecule and a dimer-based network adsorbed on Ag(111) surface, the DFT optimized dipole moment orientations seem to be slightly tilted with respect to the film plane[63] with a non-zero out of plane component of polarization. This can be used to probe the ferroelectric polarization with surface probe techniques, provided that a detailed theoretical understanding of the polarization components and their impact on a possible out-of-plane component of hydrogen transfer mechanism can be obtained a priori. Moreover, as STM is evolving to be a very effective tool for characterizing 2D ferroelectric materials, even with in-plane ferroelectric polarization[104], similar ideas can be employed to probe ferroelectricity in ultra-thin films of CA. With sufficient spatial resolution, STM may also be employed to directly image the hydrogen atoms in a surface supported network of CA molecules and track the hydrogen transfer mechanism associated with the polarization reversal.

3.3.2. Polycrystalline thin films
3.3.2.1. Growth optimization studies

Studies on the growth and ferroelectric properties of polycrystalline thin films of thickness in the range of several 10s of nanometers have been performed. These attempts are directed toward the optimization of growth conditions of CA films that maintain ferroelectricity observed in bulk crystals so as to realize the fabrication of nanoelectronic and spintronic devices. Usually, when organic materials are vacuum-grown onto substrates, two parameters, namely, the molecule-molecule interaction and the substrate-molecule interaction, determine the growth behavior of the molecular layer. The molecule-molecule interaction tends to coagulate the molecules together and encourages a three-dimensional layer growth, whereas stronger substrate-molecule interactions encourages two-dimensional growth. The three-dimensional growth mode may result in a relatively rougher surface, which impedes the growth of high-quality interfaces in vertical devices and would result in poor device performance. Thus, a major motivation behind the growth optimization studies for any organic material is to optimize the growth conditions to obtain thin two-dimensional layers with a minimum surface roughness.

As an example, CA film was grown by Jiang et al. on $Al_2O_3$ surface by high-vacuum physical vapor deposition methods at optimized substrate temperatures and thickness ranges to obtain continuous two-dimensional films of several 10s of nanometers.[47] The surface coverage of the films is highly dependent on the above two parameters and there exists an optimized range between both parameters where continuous films can be obtained. A minimum thickness of 20 nm was necessary to prepare a reasonably continuous layer. The films are grown at specified substrate temperatures to limit the surface diffusion, and thus obtained quasi-continuous films. The films were then slowly annealed up to room temperature to allow the crystallization to take place.[47] It was observed that the films are continuous only above a certain film thickness (nominal thickness of 20 nm) and for a particular range of substrate temperature (around -33º C or 240 K). However, the surface of the continuous films was not very smooth, with an *rms* roughness reaching 3.5 nm for a 5×5 µm$^2$ area. Such a roughness may be acceptable while fabricating thicker films of thickness reaching 100s of nanometers but for thinner films, it will result in poor quality interfaces and greatly hamper the performance of vertical devices.

Piezoresponse Force Microscopy (PFM) imaging of the aforementioned CA films at room-temperature revealed a polydomain structure of the samples. The grain-to-domain correspondence indicated that each grain could be a single crystalline domain of CA. Further, local polarization reversal attempts by applying electric field with the PFM tip resulted in polarization hysteresis loops with a coercive field of ~7 V, thereby confirming ferroelectricity in these films. Besides, macroscopic polarization hysteresis loops obtained from capacitor devices with Al electrodes and a 185 nm-thick dielectric CA layer resulted in a macroscopic polarization of ~0.4 µC/cm$^2$, which is much lower than what is observed in CA crystals.

It must be noted that, in a polycrystalline sample, the polarization vector in the individual grains may be oriented along random directions, with varying out-of-plane and in-plane components of polarizations. The presence of a non-zero out-of-plane component of grains in such films, contrary to the highly in-plane polarization orientation in ultra-thin films, makes it possible to study polarization reversal processes via a vertical electric field using standard microscopy probes.

From the point of view of fabricating nanoelectronic devices with ultra-thin CA films, the highly in-plane polarization on metal substrates stands as a major obstacle, as it would be difficult to control an in-plane ferroelectric polarization state with an out-of-plane electric field in a vertically structured electronic device, which is the most widely used scheme of device fabrication. On the other hand, the high surface roughness of bulk thin films, may impede the fabrication of high-quality thin film devices with CA. Furthermore, the highly anisotropic polarization in the nanoscopic grains of polydomain-polycrystalline bulk CA films may not allow the entire polarization within a macroscopic vertical device to be oriented perpendicularly to the film plane, thereby reducing the effective macroscopic polarization along the controllable direction. This calls for advanced growth methods based on physical vapor deposition that are compatible with large-scale ultra-clean device manufacturing setups and can ensure a smoother surface, a better surface coverage for ultra-thin films, robust ferroelectric properties of the deposited molecules and control over the polarization axis of the organic ferroelectric.[105]

One attempt at growth optimization is to apply a strong in-situ electric field during the growth of these polar molecules. This can be done by applying a high voltage to a conducting mesh that covers the substrate at a certain distance from it and is placed on the molecular source-substrate path. With the substrate connected to the ground, the high voltage will create a strong electric field near the substrate surface that orients the polar molecules during film growth. Parameters such as the magnitude and direction of applied electric field, distance between the substrate and the mesh, and porosity of the mesh can be tuned to control the growth behavior of the film.

Several groups have undertaken this technique. For example, by applying an electric field of 4 kV/cm to the mesh, Costa et al.[106] were able to reduce the surface roughness of a 22 nm CA film grown on Si/SiO$_2$ substrate to 1 nm (full width at half maximum in the height distribution). More importantly, the ferroelectric properties of the films grown in such a manner were shown to be retained. However, similar optimization attempts on thicker or thinner films have not been reported yet. Likewise, Yuan et al[107] studied the temperature dependence of the nucleation rate of CA films with and without electric field-assisted growth and observed that the maxima of the grain density shifts in temperature when an electric field is applied during growth. Similarly, the void defects arising at the interface between CA films and substrates such as Si and SiO$_2$ have been studied by Peterson et al.[108] Further, Hu et al.[70] have been able to grow CA films with crystallites uniaxially oriented, c-axes perpendicular to the film plane, without using any in-situ electric field and have demonstrated stable ferroelectric properties as well as non-volatile memory switching behavior stabilized by space charges.

   3.3.2.2. Toward applications of Croconic Acid thin films

Parallel to the efforts on the growth optimization of CA thin films, attempts have been made toward exploring its potential as a candidate for possible future applications, for example in the field of nanoelectronics and spintronics. For example, metal-ferroelectric-insulator-semiconductor diodes with CA as the ferroelectric have been studied by Ohmi et al.[68] Using CA as the gate, pentacene-based organic field effect transistors (OFET) have been realized, and a hysteretic behavior of the drain current versus gate voltage has been observed owing, presumably, to ferroelectricity in the gate material.[109]

Mohapatra et al.[46] have studied the growth behavior and stability of ferroelectric properties of CA thin films on ferromagnetic (for example, Co) and non-magnetic metallic (for example, Au) substrates, which are relevant as electrode materials in spintronics.[110,111] Apart from showing that the ferroelectric properties of CA thin films may depend on the type of substrate material used, on Au surface, the inability to reversibly switch the ferroelectric domains indicates the substrate surface sensitivity of ferroelectric polarization reversal properties of CA. Nevertheless, the observed multi-domain in-plane structures (figure 13a-c) hint at the 180⁰ or anti-parallel alignment of neighboring domains. On the other hand, on Co surface, robust ferroelectric switching behavior at ambient temperature and pressure was demonstrated (figure 13d).

Further, Mohapatra et al.[112] employed a unique combination of measurements to detect small polarization switching currents (a few pA) and extract a nanoscopic understanding of the ferroelectric and polarization reversal properties toward nanoscale device applications. A polarization switching current from a laterally nanoscopic region of CA thin film could be detected and the nanoscopic polarization densities could thus be estimated. Figures 14a, b show typical polarization switching current plot (a) along with the concurrent nanoscopic piezoelectric

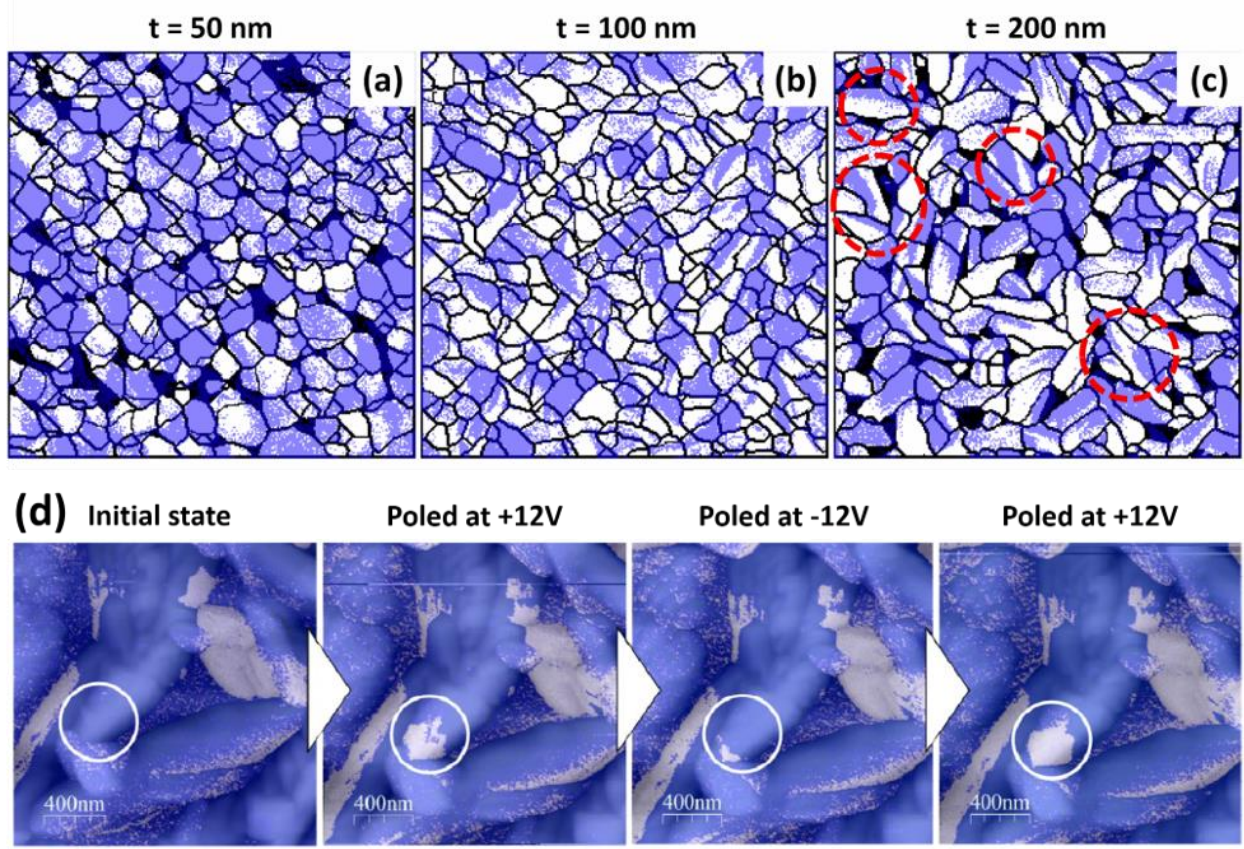

**Figure 13.** Thickness-dependent domain imaging on polycrystalline CA grown on Au surface (a to c) and robust ferroelectric polarization reversibility of CA domain on Co surface (d). In-plane domain images (white-purple contrast) superposed on the morphological map (grain boundary black lines) for three different thickness of CA films grown on Au surface are shown in (a)-(c). Some of the grains harboring multi-domain structures are indicated with dashed red lines in (c). In (d), grey-blue contrast shows in-plane domains in the LPFM phase maps of a 2×2 μm² region in the initial state, after consecutive poling at +12 V, -12 V and at +12 V, respectively. Figures are adapted from reference 46, with permission from Royal Society of Chemistry, Copyright (2020) (http://creativecommons.org/licenses/by/3.0/).

(electromechanical) strain response (b) from the studied region. The synchronization between both the features proves that the detected currents indeed originate due to true polarization reversal events in the nanoscopic region. An estimated saturation polarization value of ~7 µC/cm$^2$ has been reported from the above measurements which is much higher than the value of ~0.4 µC/cm$^2$ observed by Jiang et al.[47] and qualitatively agrees well with the polarization values reported for CA crystals. This nanoscopic detection of switching currents can provide deeper insights into the fundamental processes associated with the polarization reversal and pave the way toward exploring the kinetics of polarization reversal processes in CA thin films.[113]

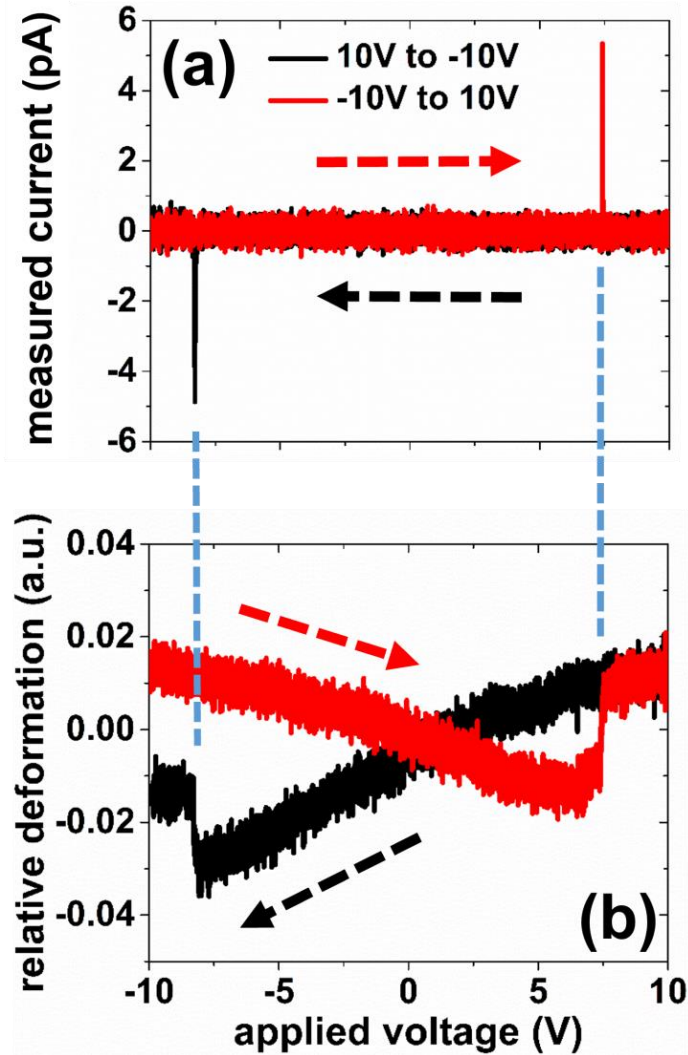

**Figure 14.** (a) and (b): measured polarization switching current from a nanoscopic region of CA thin film on Co surface, displaying the results of polarization switching current and the piezoelectric strain response measurements, respectively. Red and black curves represent the measurements for the direction of the applied voltage from -10 V to +10V (dashed red arrows) and from +10 V to -10 V (dashed black arrows), respectively. The occurrence of the current peaks is synchronous (dashed blue lines) with the jumps in the piezoelectric strain response. This establishes the veracity of the detected polarization switching current peaks.

Furthermore, Yin et al.[69] have taken the step toward integrating CA thin films into a spin valve device as a ferroelectric spacer and demonstrate spintronic applications of CA. With a 6 nm thick film of CA and 1 nm $SiO_2$ as the spacer and with Co and Lanthanum strontium manganite (LSMO) as the magnetic electrodes, they have shown the voltage control of the spin transport properties, such as magnetoresistance (MR), of the spin valve device and shown spin rectification occurring at the Co/CA interface. A bistable state of the device resistance with respect to the poling voltage for CA was observed in the device, with the low and high resistance states being called as ON and OFF states, respectively. It was observed that the congruence between the signs of the MRs in the ON and OFF states were dependent on the measuring voltage; same signs (negative for both states) at positive measuring voltage and opposite signs (negative for ON and positive for OFF state) at negative measuring voltage. These observations were the result of, and therefore signify the importance of, voltage tuning of the interfacial band alignments via the electrostatic effects of ferroelectric polarization.[114–116]

Similarly, along a different direction of work, in an attempt to create a multifunctional molecular device, Hao et al.[117] demonstrated non-volatile voltage control of the resistance across a planar device with a spin-crossover molecular layer working as the spacer. The resistance modulation is believed to originate from the bi-stable resistance states associated with the two spin states of the spin-crossover system and the role of ferroelectric CA layer is to provide the voltage control via a possible coupling between its ferroelectric polarization and the spin state of the spin-crossover layer. Although the change in resistance is not appreciably large, presumably due to the rough interface between the two molecular layers, it is highly reproducible, which hints at the potential of CA in designing multifunctional organic molecular devices.

### 4. Prospective and Outlook

This review of research on croconic acid summarizes the progress made in the past and underscores the challenges present for its future applications. First, from the point of view of application in electronic devices, the robust ferroelectricity obtained in polycrystalline films is encouraging and can be used in nanostructured ferroelectric devices. However, further efforts on growth optimization of thinner CA thin films on various substrate surfaces need to be performed, following which vertical nanostructured devices can be fabricated. Moreover, a robust control of the polarization in ultra-thin CA films on ferromagnetic surfaces can lead to the design of organic ferroelectric-based artificial multiferroic systems, which in turn can pave the way toward fabricating Organic Multi-Ferroelectric Tunnel Junction (OMFTJ) devices[84] with promising applications in spintronics. Although similar ideas have been explored with ferroelectric oxides[118] and polymers,[83],[7] the higher polarization strength and vacuum-compatible growth methods of CA films can bring crucial improvements to the device properties. Further, if ultrafast optical control of ferroelectric polarization can be achieved, CA may find substantial optoelectronic applications.

On a more fundamental side, dynamical properties of polarization reversal can be investigated at the nanoscale by measuring polarization switching currents in relation to the origin and propagation of polarization switching processes in both ultra-thin and polycrystalline thin films of CA. Well-known models[78,113,119–124] and experiments[125],[126] on ferroelectric domain and domain wall growth kinetics may be helpful for such studies. These studies can also be extended to macroscopic crystals by measuring the macroscopic polarization switching currents. Furthermore, owing to the difference in structural arrangements, it may be interesting to perform a comparative study of the reversal kinetics, in different structures such as thin films and macroscopic crystals of CA. Additionally, it will be interesting to explore the nature of ferroelectricity in the ultra-thin films of CA on metal substrates, provided, delicate characterization techniques can be designed.

With growing interest of the scientific community in the field of organic ferroelectrics, croconic acid-like, hydrogen-bonded ferroelectric systems are currently at the center of attention. Owing to ultrafast cooperative ionic and electronic mechanisms of ferroelectric polarization switching at small electric fields, hydrogen-bonded systems offer the possibility of designing energy-efficient electronic devices. Fueled by the discovery of robust ferroelectricity in CA, several other hydrogen-bonded room-temperature ferroelectric molecular systems, such as 2-phenylmalondialdehyde (PhMDA), 3-hydroxyphenalenone (HPLN), cyclobutene-1,2-dicarboxylic acid (CBDC), methylbenzimidazole (MBI), 5,6-dichloro-2-methylbenzimidazole (DC-MBI)[127],[128],[43] and 3-anilinoacrolein anil (ALAA)[129],[43], have been discovered. These systems also are known as proton tautomerism-based hydrogen-bonded ferroelectric systems. Although CA leads the category with the largest polarization value, insights may be obtained from its ferroelectric properties toward designing additional hydrogen-bonded or proton transfer-based systems with an even larger spontaneous polarization.

Evidently, the future directions of research on croconic acid involve and combine different subject areas, starting from chemistry to material science. Thus, thanks to an interdisciplinary focus, the future potential of this relatively new organic ferroelectric can be unraveled. We believe this review pertains to a broad area of research on croconic acid and will work toward encouraging scientific communities to further explore this material.

**Acknowledgements:**

We are thankful to the funding agency ANR for supporting the project ORINSPIN (ANR-16-CE92-005-01). This work is a tribute to the late Dr Eric Beaurepaire of IPCMS, CNRS and we respect his efforts in designing the project involving croconic acid.